\documentclass{ws-p8-50x6-00}
\usepackage{epsfig}

\def\jour#1#2#3#4{{#1} {\bf#2}, #4 (19#3)}

\def\PL{{\em Phys. Lett.}  {B}}   

\def\EPJ{{\em Eur. Phys. J.} {C}}
\def\NPB{{\em Nucl. Phys.} {B}}
\def\PRp{\em Phys. Reports}

\def\IJ{{\em Int. J. Mod. Phys.} {A}}

\def\vep{\varepsilon}
\def\vtt{\vartheta}

\def\be{\begin{equation}}
\def\ee{\end{equation}}
\def\bea{\begin{eqnarray}}
\def\eea{\end{eqnarray}}
\def\bc{\begin{center}}
\def\ec{\end{center}}
\def\bi{\bibitem}
\def\lb{\label}
\def\ct{\cite}
\def\ea{\sl et al.}

\def\ifmath#1{\relax\ifmmode #1\else $#1$\fi}%

\def\rt{\ifmath{{\mathrm{t}}}}
\def\vec#1{{\mbox{\bf #1}}}

\begin{document}

\title{Intermittency and Correlations\\ at LEP and at HERA}

\author{EDWARD K. G. SARKISYAN}

\address{EP Divison, CERN, CH-1211 Geneva 23, Switzerland
\\and\\
HEP Department, School of Physics and Astronomy, Tel Aviv University,\\
IL-69978 Tel Aviv, Israel 
\\ 
E-mail: Edward.Sarkisyan@cern.ch}


\maketitle

\abstracts{
A review on recent investigations of local fluctuations and genuine
correlations in e$^+$e$^-$ annihilations at LEP and in e$^+$p collisions 
at HERA is given. 
}
\vspace*{-0.9cm}

\section{Introduction}\lb{sec:intro}
\vspace*{-0.2cm}

Local fluctuations and genuine correlations of hadrons in high energy
interactions provide us with important information about the multihadron
production mechanism. This information gives detailes which are beyond
those obtained from studies of single-particle distributions. 

The peculiarity of hadrons to group when produced have been
studied a long time using correlation functions, while in the recent
decade the
interest in correlations have been revived due to a new method of
factorial moments and due to the obtained intermittency
phenomenon, i.e. self-similarity, or fractality, of
hadron production.\ct{rev1,rev2}

The fractal structure of particle distribution in e$^+$e$^-$ collisions
has been
realized many years ago due to a jet evolution picture (see e.g.
\ct{rev2}). 
However,
even if the
parton shower is
expected to exhibit intermittency, this does not guarantee the effect to
appear at hadron level. Recently, different approaches have been
proposed
to describe the experimentally observed fractality by analytical QCD
calculations.\ct{revo} 

In this talk, we review experimental studies of intermittency and
correlations at LEP \ct{l3a,l3f,da2,da12,opf}
and at HERA,\ct{ha2,ha} which appeared recently and provide us
with further information, in addition to  discussions of recent 
reviews.\ct{rev1,rev2} 
From the studies, it is seen that, despite considerable success in
understanding different properties of the phenomenon, further
investigations are
needed.  

\section{Definitions}\lb{sec:defs}
\subsection{A Tool of Factorial Moments and Cumulants}
\lb{subsec:tool}

In order to measure local dynamical fluctuations, a method of
normalized factorial moments, $F_q$, is applied.\ct{rev1,rev2} 
The factorial
moment of
order $q$  is defined as a function  of a phase-space region size
$\delta$,

{  \vspace*{-0.37cm}
\be
F_q(\delta)=\langle n(n-1)\cdots(n-q+1)\rangle /\langle n\rangle ^q.
\lb{eq:fq}
\ee
  \vspace*{-0.4cm}
}
  
\noindent 
Here, $n$ is the number of particles in the $\delta$-region, and 
the brackets $\langle \cdots \rangle$ denote averaging over events. 

The normalised factorial moments allow us to extract dynamical
fluctuations. For uncorrelated particle production, the moments
are independent of
$\delta$, $F_q\equiv 1$.   
Correlations beetween particles lead to an increase of factorial moments with
decreasing $\delta$ (increasing number $M$ of $\delta$-regions),  and
if exhibiting a
power-law the dependence is called intermittency.

To extract genuine correlations contributing to the fluctuations,
one uses
the technique of normalised factorial cumulant moments,
cumulants.\ct{rev1,rev2}
The  cumulants, $K_q$, are constructed from the unnormalised factorial
moments in a way that they vanish whenever particles in
$q$-tuple are statistically independent.
The normalised cumulants are defined as

{  \vspace*{-0.4cm}
\be
K_q(\delta)=k_q/\langle n\rangle ^q.
\lb{eq:kq}
  \vspace*{-0.2cm}
\ee
  \vspace*{-0.2cm}
}

\noindent with the Mueller moments $k_q$, 
\bea
\nonumber
k_1=\langle n\rangle \:, \quad k_2=\langle n(n-1)\rangle - \langle
n\rangle
^2,\\
k_3=\langle n(n-1)(n-2)\rangle -3\,\langle n(n-1)\rangle\,\langle n\rangle
+2\,\langle n\rangle ^3\, , \quad {\rm etc.}
\lb{eq:km}
\eea
Normalised cumulants share with the normalised factorial moments their
property to measure the dynamical
component of the underlying particle density. 

From Eqs. (\ref{eq:kq}) and (\ref{eq:km}), one finds the interrelations
between normalised factorial moments and cumulants,

{  \vspace*{-0.5cm}
\be
F_1 = K_1, \quad
F_2   =  K_2 + 1, \quad
F_3  =  K_3 + 3K_2+ 1,  \lb{eq:kf} \; {\rm etc.}\ ,
\ee
\nonumber
\vspace*{-0.42cm}
}

\noindent which provide us with the information whether the
$p$-order genuine correlations are important in the $q$-particle dynamical
fluctuation.

\subsection{Analytical QCD Predictions}\lb{subsec:QCD}

The QCD description of multihadron production is based on a partonic
picture, i.e. on gluon cascades radiated off the initial parton.\ct{revo}
In order to describe the hadron production mechanism, one has to cut off 
the parton cascade
at some scale $Q_0\leq 1$ GeV, while the following non-perturbative
hadronization is considered within the concept of Local Hadron Parton
Duality (LPHD), which connects multihadron final states and 
partons.

The QCD calculations are given in angular phase space, i.e. in
1-dimensional, 1D, rings (or 2-dimensional, 2D, cones)  around a jet axis
with
mean opening angle $\Theta$ (a direction ($\Theta, \Phi$)) and a half
width
(opening angle) $\delta \equiv \vtt$. 

For the normalised cumulants and factorial moments, the power-law,

{  \vspace*{-0.3cm}
\be
\nonumber
K_q(\Theta, \vtt) \:\: {\rm or} \:\: F_q(\Theta, \vtt) \propto
(\Theta/\vtt)^{(q-1)(D-D_q)},
\lb{eq:sq}
\ee
\vspace*{-0.4cm}
}

\noindent with fractal or R\'enyi dimensions $D_q$ is predicted.    $D$
is a
dimensional factor: $D=1$ for ring regions and 2 for cones.

The QCD expectations for $D_q$ are as follows (see \ct{revo} and refs.
therein).

\begin{itemize}

{ \vspace*{-0.2cm}
\item In a fixed-coupling regime ($\alpha_s = {\rm const.}(\vtt)$) of
the Double Log Approximation (DLA),
} 
{  \vspace*{-0.2cm}
\be
D_q\equiv D_q^{\rm (c)}=\gamma_0(Q)\, \frac{q+1}{q}, \quad
\gamma_0(Q)=\sqrt{2N_c\alpha_s/\pi}, 
\quad Q=E\Theta,
\lb{eq:dc}
\ee
}
for  moderately small angular regions, $\vtt\leq \Theta$.

\item In a running-coupling regime of the DLA, 
{  \vspace*{-0.1cm}
$$
\hspace*{-0.75cm}
{\rm (a)} \quad D_q=D_q^{\rm (c)}\left(1+
\frac{q^2+1}{4q^2}\cdot \vep \right),\quad
\:
{\rm (b)}\quad
D_q=2D_q^{\rm (c)}
\left(
\frac{1-\sqrt{1-\vep}}\vep \right),
$$ 
}
{  \vspace*{-0.2cm}
\be
{\hspace*{-1.37cm}
{\rm (c)}\quad
D_q=2\gamma_0(Q)\,
\frac{q-w(q,\vep)}{\vep(q-1)}}, \quad
w(q,\vep)=q\sqrt{1-\vep}
\left(1-
\frac{\ln(1-\vep)}{2q^2} \right).
\lb{eq:dq}
\ee
}
{  \vspace*{-0.4cm}
\item In the Modified Leading Log Approximation (MLLA), Eq. (\ref{eq:dq}a)
remains
valid but $\gamma_0$ is replaced by an effective  one, $\gamma_0^{\rm
eff}(Q)=\gamma_0(Q)+\gamma_0^2(Q)\cdot f(q,N_f,N_c)$.}  
\vspace*{-0.2cm}
\end{itemize} 

\noindent 
Here, $E$ is the jet energy, $N_c$ and $N_f$ are the number of 
colors and flavors, respectively. 

A scaling variable,   

{\vspace*{-0.4cm}
\be
\vep = \frac
{\ln(\Theta/\vtt)}
{\ln(E\Theta/\Lambda)}\, 
\lb{eq:ez}
\vspace*{-0.2cm}
\ee
}

\noindent is utilised in the calculations.
For the maximum phase space, $\vtt=\Theta$, $\vep=0$.

The analytical predictions involve only one adjustable parameter, the QCD
scale $\Lambda$, while a strong coupling $\alpha_s$ is based on
first-order QCD relation,

{\vspace*{-0.5cm}
\be
\alpha_s = {{\pi\beta ^2}\over{6}}{1\over{\ln(Q/\Lambda)}}, \quad
\beta^2=12 \left({11\over{3}}N_c-{2\over{3}}N_f\right)^{-1}.
\lb{eq:fo}
\ee
\vspace*{-0.2cm}
}

The QCD calculations are made at
asymptotic energies, which corresponds to an infinite number of partons in
an event. No energy-momentum conservation is taken into
account in the calculations above.  
%%%\vspace*{-2.cm}

%#01%
\begin{figure*}[!b]
\vspace*{-.45cm}
\epsfysize=6.4cm
\epsffile[23 381 529 644]{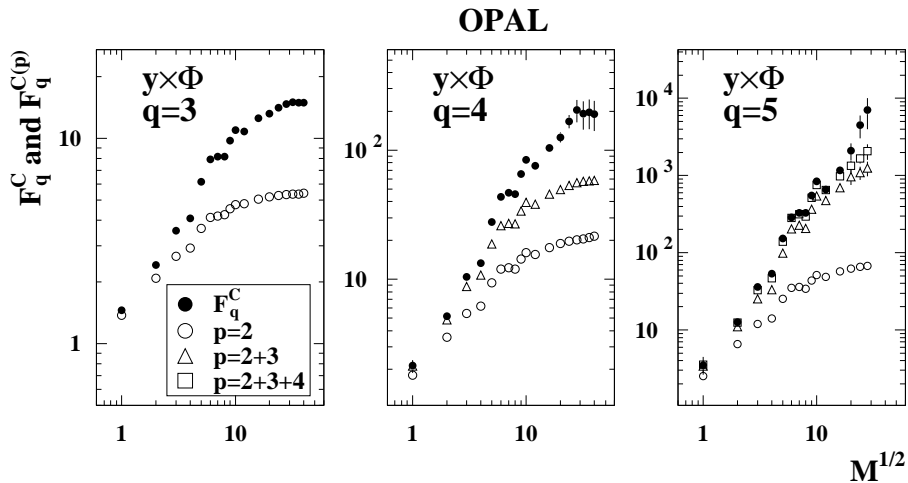}
\vspace*{-.8cm}
\caption{Decomposition of factorial moments $F_q$ into
correlation contributions $F_q^{(p)}$ in the subspace of rapidity 
vs. azimuthal angle, measured by OPAL $^8$  (see Eq. (\ref{eq:kf}) and
text).
}
\lb{fig:fk}
\vspace*{-.5cm}
\end{figure*}

\section{Experimental Results}\lb{sec:res}
\vspace*{-.2cm}

\subsection{Spatial Fluctuations and Correlations}
\vspace*{-.05cm}

At LEP, L3 \ct{l3f} and OPAL \ct{opf} have studied fluctuations and
correlations in the Z$^0$$\to$e$^+$e$^-$$\to $hadrons process. 

L3 measured fluctuaions in rapidity and in the 4-momentum difference
($Q^2_{ij}=-(p_i-p_j)^2$ of all $ij$-pairs)  using factorial moments for
the former variable and the bunching parameters for both variables. 
The measurements have shown the multifractal character of the local
fluctuations, as it is expected from the QCD parton shower picture.

A large statistics of more than 4 million events has been used by OPAL to
measure local fluctuations and genuine correlations in one-, two- and
three-dimensional subspaces of
rapidity, azimuthal angle, and transverse momentum (w.r.t. the sphericity
axis).
As L3,  OPAL has observed
multifractal behavior of factorial moments in one dimension, rapidity and
azimuthal angle. Such an intermittency behaviour gets more pronounced with
increasing dimension, which is ascribed to a jet-like structure of the
events as predicted.\ct{rev2} 

OPAL has measured factorial cumulants and found genuine correlations
up to 5th order, being especially large in the rapidity vs. azimuthal
angle subspace. Using Eqs.~(\ref{eq:kf}) reduced to some $p$th order,
$p$$<$$q$ (e.g. $F_3^{(2)}=3K_2+1$), OPAL checked the importance of the
genuine correlations in the
dynamical fluctuations obtained. It was found that genuine
correlations 
of high-order are needed to describe  the intermittency
effect, see Fig. \ref{fig:fk}.

In both Collaborations' studies, Monte Carlo (MC) models which
have been tuned to reproduce global event-shape distributions and
single-particle inclusive spectra in e$^+$e$^-$ annihilations, were found
to reproduce the trend, but not the magnitudes of the measured moments,
especially for
small $\delta$.
\vspace*{-.2cm}

\subsection{Angular Fluctuations and Correlations}\lb{subsec:afq}

L3 \ct{l3a}, DELPHI \ct{da2,da12} and ZEUS \ct{ha2,ha} have measured
fluctuations and correlations in angular
phase space. While the thrust or sphericity axes are chosen as a jet axis
in LEP experiments,
the special Breit frame is considered in the HERA experiment, to separate
the hadronic final state from the radiation and to mimic 
a single e$^+$e$^-$ reaction hemisphere.

%#02%
\begin{table}[!b]
\vspace*{-.35cm}
{\hspace*{-0.25cm}
\begin{tabular}{lr}
\epsfig{file=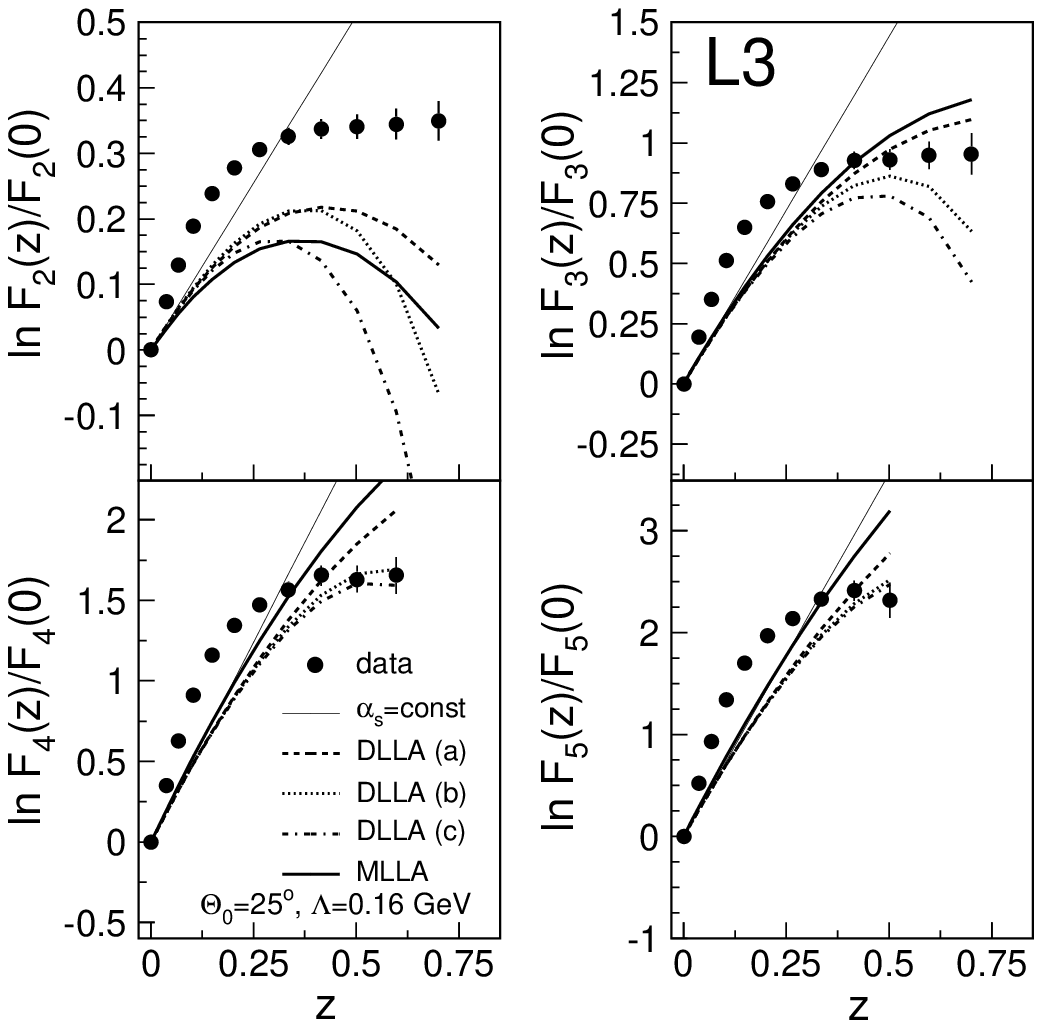,height=5.58cm}
&
\hspace*{-0.4cm}
\epsfig{file=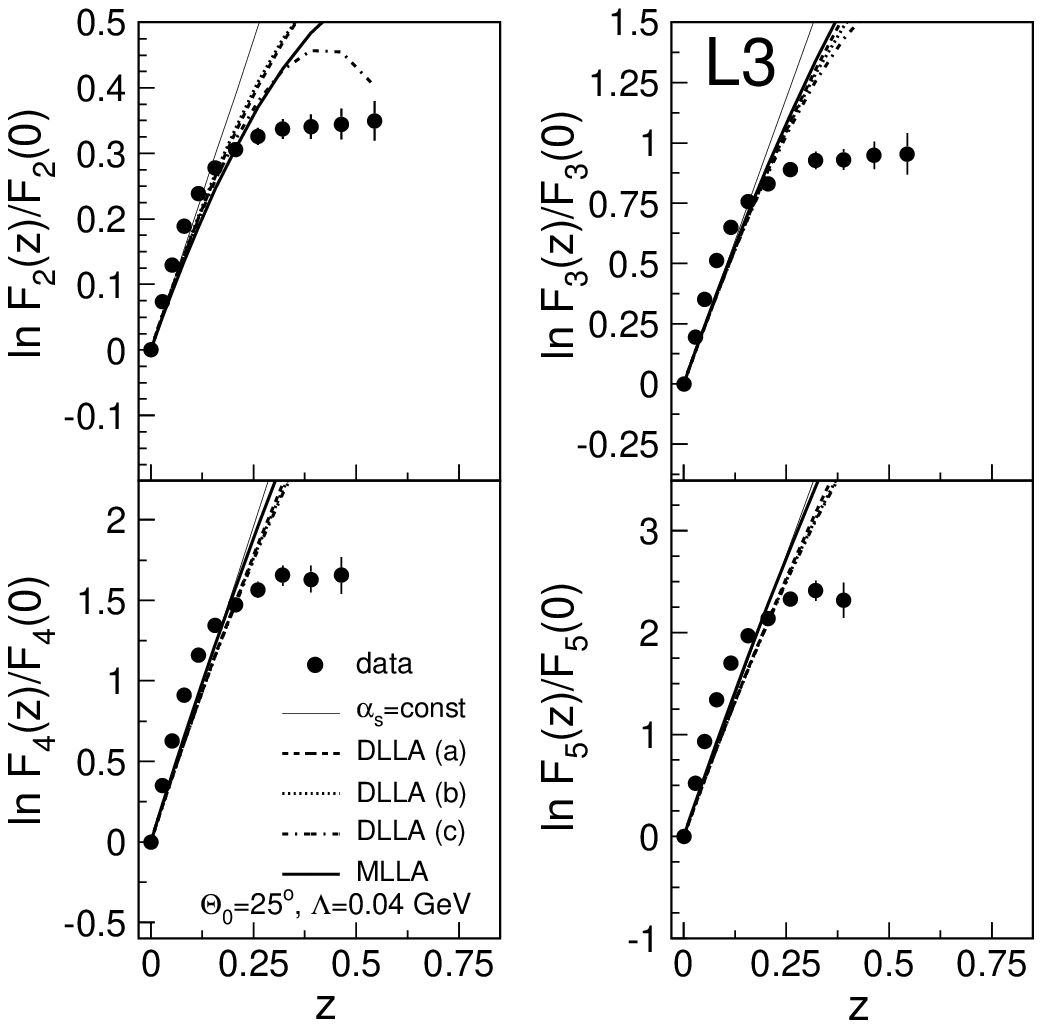,height=5.58cm.}
\end{tabular}
}
{\vspace*{-.1cm}
\begin{flushleft} {\footnotesize{Figure 2:  
L3 $^4$ factorial moments $F_q(z\equiv\vep)/F(0)$ at $\Lambda$=0.16 GeV
(left)
and $\Lambda$=0.04 GeV (right) compared to the QCD calculations according
to DLA
Eqs. 
(\ref{eq:dc}), (\ref{eq:dq}) and MLLA
($\gamma_0$ $\to\gamma_0^{\rm eff}$).}}\end{flushleft}}    
\end{table}

\setcounter{figure}{2}

%#03%
\begin{figure*}[!t]
\bc
\epsfysize=5.6cm
\epsffile[153 603 455 768]{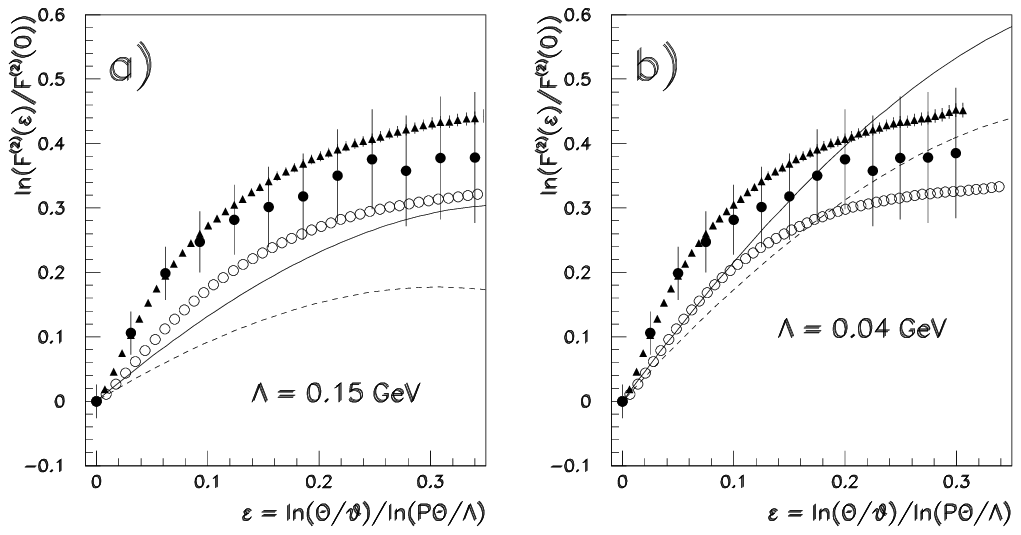}
\ec 
\vspace*{-0.65cm}
\caption{DELPHI factorial moments $F_2(\vep)/F_2(0)$ vs. QCD
(Eq.(\ref{eq:dq}c),
$N_f$=3)
and MC predictions.$^7$ Data and the QCD results
are shown by, respectively, open circles and dashed lines
at $\sqrt{s}$=91 GeV and by full circles and solid lines at   
$\sqrt{s}$=183 GeV. The full triangles denote the {\sc Jetset} MC 
results at 183 GeV.
Similar QCD predictions are claimed to be found if
one uses Eqs.  (\ref{eq:dq}a) or (\ref{eq:dq}b).  
} 
\vspace*{-0.7cm}
\lb{fig:he}
\end{figure*}

Comparing the measurements to the first-order analytical predictions,
Eqs. (\ref{eq:dc}), (\ref{eq:dq}), and (\ref{eq:fo}), it
has been found that the DLA and MLLA qualitatively describe 
the general features of the data (Figs. 2, \ref{fig:he}). The
measured moments rise
approximately linearly for large angles $\vtt$ (small $\vep$)
as expected from the parton shower multifractality (cf. Eq.
(\ref{eq:sq})),
while the levelling off at smaller angles is believed to be 
due to the runing effect of
the coupling $\alpha_s$. The 2D moments rise much
more steeply than the 1D ones.\ct{da12} The factorial
moments are found \ct{da12} to increase as the energy increases, see Fig.
\ref{fig:he}.
Note that the $\vtt$-
and D-dependences are analogous, respectively, to the multifractality of a
parton shower and  to the jet structure, obtained in spatial
analyses,\ct{l3f,opf} as discussed above.

On the quantitative level, one finds some deviations between the QCD 
predictions and the data. 
The analytical
calculations
are very sensitive to the QCD parameters, $\Lambda$ and $N_f$, and are not
able to describe simultaneously the factorial moments at all orders and at
different dimensions. A better agreement
between the factorial moments calculated and the data is obtained for
$\Lambda$=0.04 GeV than that is found at the  expected larger value,
$\Lambda$=0.16 GeV,
Figs. 2 and \ref{fig:he}. However, at small $\Lambda$,
the theory overestimates the data for large
$\vep$. 
The measured cumulants are far from the predictions, and the  
reduction of the $\Lambda$-value was not found to sensibly 
improve the situation, see Fig. \ref{fig:akq}.   
%#04%
\begin{figure*}[!t]
\vspace*{-0.3cm}
\bc
\epsfysize=7.9cm
\epsffile[123 546 514 775]{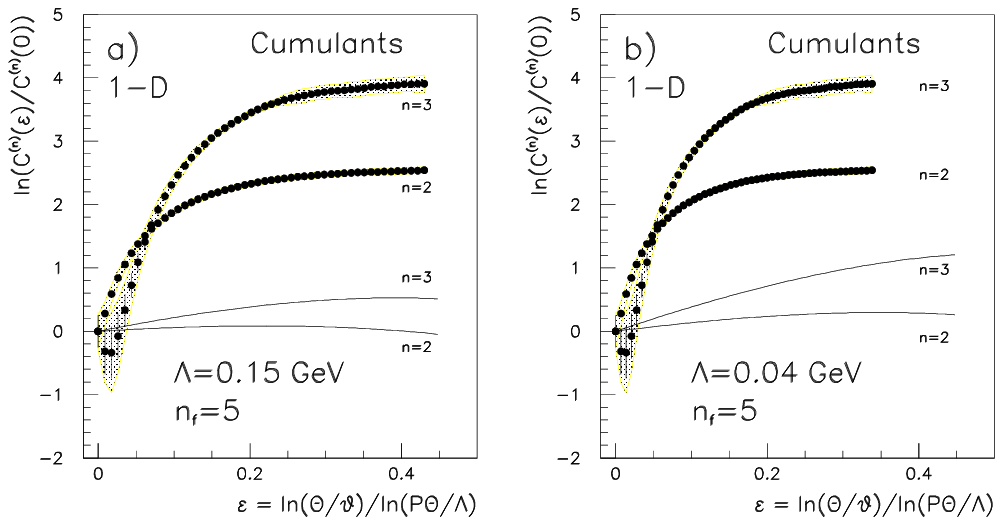}
\ec
\vspace*{-2.5cm}
\caption{DELPHI cumulants (circles), calculated via Eq. (\ref{eq:kf}), are 
compared to the QCD Eq. (\ref{eq:dq}c)
calculations (lines).$^7$ 
The statistical errors (error bars) are shown along with the systematic
ones (shaded areas).}
\lb{fig:akq}
\vspace*{-0.2cm}
\end{figure*}
%#05-06%
\begin{flushleft}
\begin{table}[!t]
\vspace*{-0.05cm}
{\hspace*{-0.25cm}
\begin{tabular}{lr}
\epsfig{file=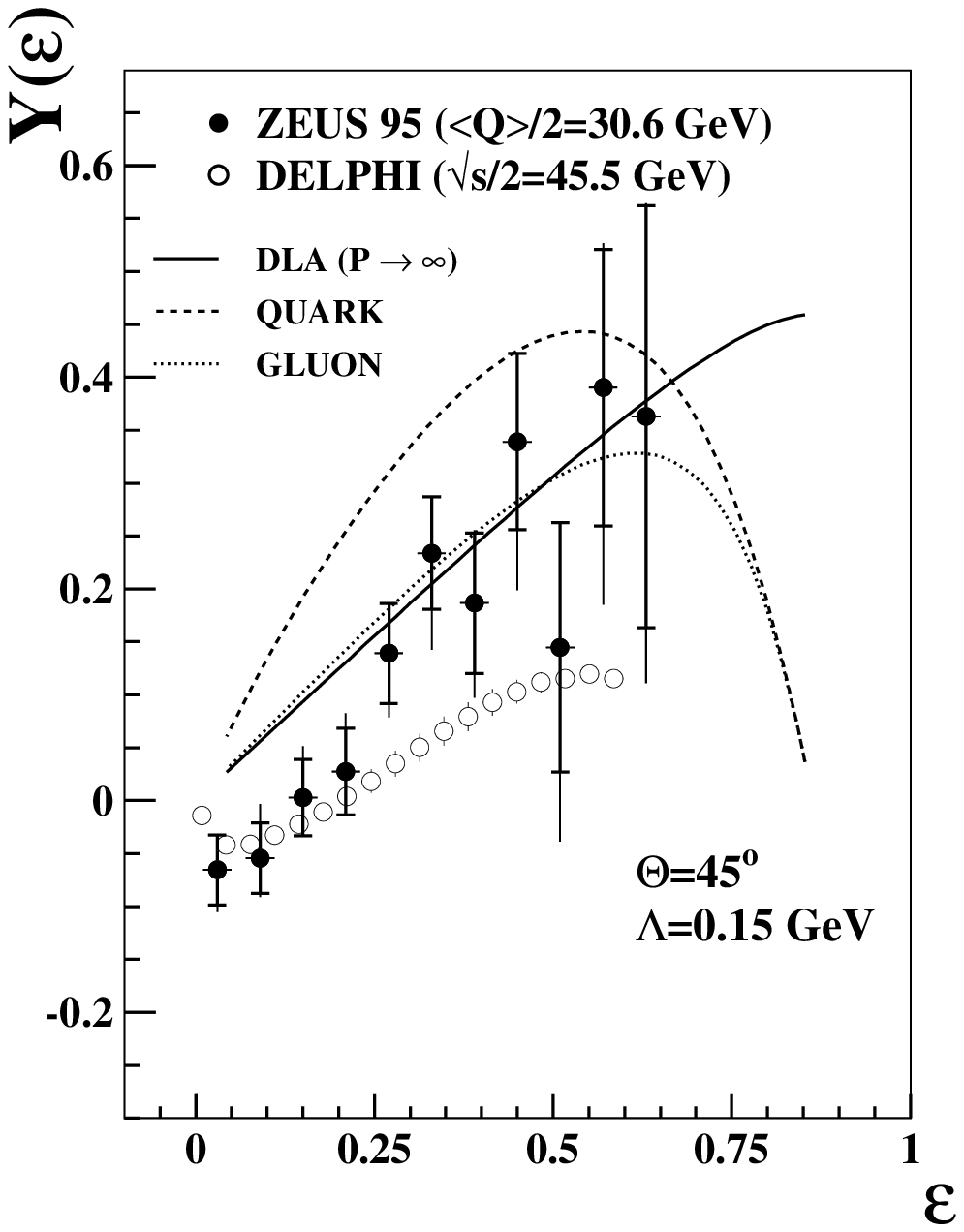,width=5.58cm,height=7.cm}
&
\hspace*{-0.7cm}
\epsfig{bb= 140 445 445 770,file=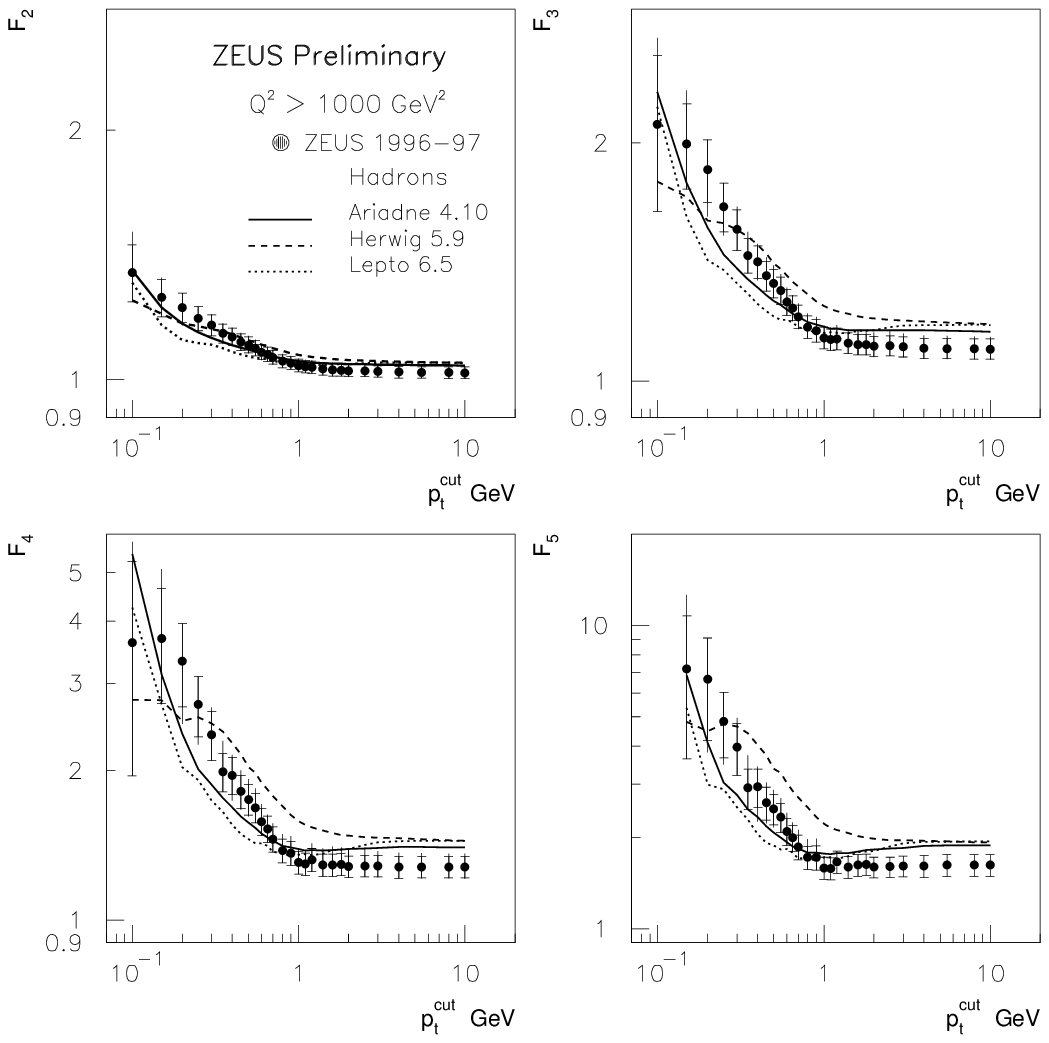,height=6.5cm.}
\end{tabular}
}
{\vspace*{-.06cm}
\begin{flushleft} {\footnotesize{
Figure 5 (left):  ZEUS 2-particle correlation function, compared to the
runing $\alpha_s$ QCD calculations (solid line), to results of DELPHI
$^6$, and to predictions for 30.6 GeV quark and gluon jets (dashed and
dotted lines, repectively).$^9$ The statistical and total errors are
shown by the inner and outer bars, respectively.\par}}

\smallskip
{\footnotesize{Figure 6 (right): ZEUS $^{10}$ factorial moments
$F_q(p_\rt^{\rm
cut})$, compared to MC models. The statistical and total errors are shown
by the inner and outer bars, respectively.\par}}
\end{flushleft}}    
\vspace*{-0.5cm}
\end{table}
\end{flushleft}

%%%\vspace*{-0.15cm}
Likely reasons for the failure of the QCD calculations seem to be their
asymptotic character and lack of energy-momentum conservation, as it was
mentioned in Sec. {\ref{subsec:QCD}. However, DELPHI has analysed some
high-energy events (Fig. {\ref{fig:he}) and no improvement at
small $\vep$-values was found.
Inclusion of energy-conservation terms was observed to 
lead to even larger discrepancies.\ct{da12} 

ZEUS \ct{ha2} has compared their results on 2-particle angular correlation
functions with the DELPHI analysis \ct{da2} to check energy-dependence of
the
correlations, see Fig. 5. According to DELPHI, there is a steeper rise 
of the 2-particle correlation functions at
$\sqrt{s}$=183 GeV than at $\sqrt{s}$=91 GeV. No such dependence is
confirmed by Fig. 5, although ZEUS data are taken at lower
energy
than those of DELPHI. All the possible checks (of experimental and
calculation procedures) kept the results unchanged. It would be
interesting to carry out an analysis to understand whether the
expected universality of the 
2-particle inclusive density is violated.
Note that QCD calculations for a gluon jet 
describe the data\ct{ha2} better than those of a quark jet,
as shown in Fig. 5. This could be connected
with the approximation used (DLA instead of, e.g., MLLA) in calculating
the ratio of the mean multiplicity in gluon and quark jets. 
Nevertheless, it is seen that both predictions disagree
with the data for small $\vep$ values (large $\vtt$).

A better agreement between the data and the calculations is found when one
compares the measurements with MC simulations. At the Z$^0$ peak, as well
as at high energies, the simulations reproduce the data well. This is
presumably due
to the fact that MC models take into account the energy-momentum
conservation
and are tuned to the data global variables. On the other hand,
there is
still a difference in  the choice of the cut-off parameter $Q_0$, which
"terminates" the parton cascade: 
$Q_0$ is about 0.3-0.6 GeV in
the MC models used, while due to the LPHD it is expected to be 
$\simeq$0.25 GeV. 
However, although  parton level MC studies by L3 indicate
some disagreement with the LPHD assumption, the DELPHI investigation tells us
that even a possible violation of LPHD seems unlikely to be a reason for the
discrepancies between hadron and parton levels.

\subsection{Fluctuations, Correlations and QCD
Coherence}\lb{subsec:coh}

Recently, ZEUS \ct{ha} has studied correlations in momentum-restricted
regions in view of recent QCD+LPHD calculations\ct{coh}. 
The normalised factorial moments of the multiplicity distributions are
theoretically expected to behave as

{\vspace*{-0.3cm}
\be
F_q(p_\rt^{\rm cut})\simeq 1+{q(q-1)\over{6}}
\frac{\ln(p_\rt^{\rm cut}/Q_0)}{\ln(E/Q_0)}\,, \qquad F_q(p^{\rm
cut})\simeq{\rm const}>1,
\lb{eq:coh}
\ee
}

\noindent when particles are restricted
(cylindrically) in either the transverse momentum $p_\rt<p_\rt^{\rm
cut}$ or (spherically) in absolute momentum $|{\: \vec{p}}\:\:|<p^{\rm
cut}$. 

The predictions of Eq. (\ref{eq:coh}) are the followings. There are
correlations between partons because the factorial moments exceed unity.
The correlations vanish for small $p_\rt^{\rm cut}$, $F_q\to 1$ as
$p_\rt^{\rm cut} \to Q_0$, due to angular ordering of the partons in the jet
(QCD coherence). This leads to the Poissonian (independent)  emission.
However, soft gluons with spherically limited momenta ($|{\:
\vec{p}}\:\:|<p^{\rm cut}$) obey the non-Poissonian distribution even for
small $p^{\rm cut}$.

Fig. 6 of 
ZEUS shows a first evident disagreement between the LPHD hypothesis and
the measurements.
The data factorial moments represent a strongly increasing function
of $p_\rt$ at $p_\rt<1$ GeV which disagrees with 
what the theory predicts, Eq.
({\ref{eq:coh}).  A similar behaviour is found for the factorial moments
in $p^{\rm cut}$, while the errors are too large to conclude. MC
models show good agreement with the data. 
 
  It would be interesting to find out whether this observation in
e$^+$p collisions agrees with that in e$^+$e$^-$ annihilation, where
one does not need to transform to a specific frame, a possible
reason of the above-mentioned disagreement in 2-particle correlation
functions. 
\vspace*{-0.3cm}

\section{Conclusions}\lb{sec:sum}
\vspace*{-0.1cm}

A review on recent results from the investigations of intermittency and
genuine correlations by DELPHI, L3 and OPAL in e$^+$e$^-$ and by
ZEUS in e$^+$p collisions is given. The
tool of factorial multiplicity and cumulant moments has been
applied. The studies
show an
existence of strong correlations between produced hadrons.  The analytical
QCD calculations qualitatively describe the scaling behaviour of the measured
moments, while on the quantitative level some discrepancies are found, 
especially for genuine correlations. Monte Carlo models
reasonably well reproduce the trend in the data, although some
disagreement in magnitudes is seen. ZEUS observed 
a violation of LPHD
predictions for momentum-limited factorial moments. The universality
between e$^+$e$^-$ and e$^+$p collisions is analysed.
The observations reviewed give clear evidence for the need of further 
efforts on studying the subject. 
\vspace*{-0.2cm}

\section*{Acknowledgments}
\vspace*{-0.2cm}
I would like to thank the Organisers of the Symposium for inviting me,
partial financial support and warm atmosphere created during the meeting.
I thank B.~Buschbeck, S.V.~Chekanov, I.M.~Dremin, W.~Kittel, and  W.~Ochs,
for their kind assistance and useful discussions.

\end{document}